\documentclass[prl,aps,twocolumn,superscriptaddress,amsmath,amssymb]{revtex4}
\usepackage{graphicx,color}

\newcommand{\beq}{\begin{equation}}
\newcommand{\eeq}{\end{equation}}
\newcommand{\bqa}{\begin{eqnarray}}
\newcommand{\eqa}{\end{eqnarray}}

\newcommand{\dg}{^\dagger}

\newcommand{\erf}[1]{Eq.~(\ref{#1})}

\newcommand{\erfand}[2]{Eqs.~(\ref{#1}) and (\ref{#2})}

\newcommand{\bra}[1]{\left\langle{#1}\right|}
\newcommand{\ket}[1]{\left|{#1}\right\rangle}

\newcommand{\sch}{Schr\"odinger}

\newcommand{\cu}[1]{\left\{ {#1} \right\}}
\newcommand{\ro}[1]{\left( {#1} \right)}

\newcommand{\st}[1]{\left| {#1} \right|}

\newcommand{\tp}{^{\top}}

\newcommand{\blk}{\color{black}}

\renewcommand{\section}[1]{{\em #1}. --- }
\renewcommand{\hat}{}
\newcommand{\paper}{Letter}

\newcommand{\xfrac}[2]{{#1}/{#2}}

\newtheorem{theorem}{Theorem}

\newtheorem{lemma}[theorem]{Lemma}

\newenvironment{proof}[1][Proof]{\noindent\textbf{#1.} }{\ \rule{0.5em}{0.5em}}

\begin{document}

\title{Steering, Entanglement, Nonlocality, and the EPR Paradox}

\author{H. M. Wiseman}
\affiliation{Centre for Quantum Computer Technology, Centre for Quantum Dynamics,
Griffith University, Brisbane 4111 Australia}

\author{S. J. Jones}
\affiliation{Centre for Quantum Computer Technology, Centre for Quantum Dynamics,
Griffith University, Brisbane 4111 Australia}

\author{A.\ C.\ Doherty} \affiliation{School of Physical Sciences,
  University of Queensland, Brisbane 4072 Australia}


\begin{abstract}
The concept of steering was introduced by \sch\ in 1935 as a
generalization of the EPR paradox for arbitrary pure bipartite
entangled states and arbitrary measurements by one party. Until now,
it has never been rigorously defined, so  it has not been known (for
example) what mixed states are steerable (that is, can be used to
exhibit steering). We provide an operational definition, from which we
prove (by considering Werner states and Isotropic states)
that steerable states are a strict subset of the entangled states,
and a strict superset of the states that can exhibit Bell-nonlocality.
For arbitrary bipartite Gaussian states we derive a linear matrix inequality
that decides the question of steerability via Gaussian measurements,
and we relate this to the original EPR paradox.
\end{abstract}


\maketitle


The nonlocality of entangled states, a key feature of quantum
mechanics (QM), was first pointed out in 1935 by Einstein
Podolsky and Rosen (EPR) \cite{EinEtalPR35}. They considered a
general non-factorizable pure state of two systems, held by two
parties (say Bob and Alice): \beq \label{psient} \ket{\Psi} =
\sum_{n=1}^{\infty} c_n\ket{\psi_n}\ket{u_n} = \sum_{n=1}^{\infty}
d_n\ket{\varphi_n}\ket{v_n}, \eeq where $\cu{ \ket{u_n}}$ and $\cu{
\ket{v_n}}$ are two orthonormal bases for Alice's system.
If Alice chose to measure in the $\cu{\ket{u_n}}$ ($\cu{\ket{v_n}}$)
basis, then  she could instantaneously project Bob's system  into
one of the states $\ket{\psi_n}$ ($\ket{\varphi_n}$). For EPR, the
fact that the ensemble of $\ket{\psi_n}$s is different from the
ensemble of $\ket{\varphi_n}$s was problematic because ``the two systems no
longer interact, [so] no real change can take place in [Bob's]
system in consequence of anything that may be done to [Alice's]
system.'' Thus, they thought (wrongly) that this nonlocality must
be an artefact of the {\em incompleteness} of QM.
This intuition was supported by their famous example (the EPR ``paradox'') involving
position and momentum, which could be trivially resolved by considering
local hidden variables (LHVs) for $q$ and $p$.

The EPR paper provoked an interesting response from \sch\   
\cite{SchPCP35}, who introduced the terms {\em
entangled state} for states of the form of \erf{psient} and {\em
steering} for Alice's ability to affect Bob's state through her
choice of measurement basis.
\sch\ had invented the quantum state $\psi$ for atoms
\cite{SchAdP26}, and, {\em unlike} EPR, believed it gave a complete
and correct description for a localized, isolated system. Thus, he
rejected their suggested explanation of steering in terms of LHVs.
However, {\em like} EPR, he could not easily accept nonlocality, and
so  suggested that QM was {\em incorrect} in its description of
delocalized entangled systems \cite{SchPCP35}. That is, he thought
(wrongly) that Bob's system has a definite state, even if it is
unknown to him, so that steering would never be seen experimentally.
We call this a local hidden state (LHS) model for Bob.

In this \paper\ we revisit \sch's concept of steering  (which has
received increasing attention in recent years
\cite{VujHerJPA88,VerPhD02,CliEtalFoP03,SpePRA07,KirFPL06}) 
from a quantum information perspective. 
That is, we {\em define} it according to a {\em task}.

First, let us define the more familiar concept of Bell-nonlocality
\cite{BelPHY64} as a task, in this case with three parties. Alice
and Bob can prepare a shared bipartite state, and repeat this any
number of times. Each time, they measure their respective parts.
Except for the preparation step, communication between them is
forbidden. Their task is to convince Charlie (with whom they can
communicate) that  the state they can prepare is entangled. Charlie
accepts QM as correct, but trusts neither Alice nor Bob. If the correlations {\em can} be explained by a LHV model then
Charlie will not be convinced that the state is entangled; the
results could have been fabricated from shared classical randomness. Conversely, if the
correlations between the results they report {\em cannot} be so explained, then the state must be entangled.
Therefore they will succeed in their task iff (if and only if) they
can demonstrate Bell-nonlocality.

\blk
The analogous definition for steering 
uses a task with two parties.
Alice can prepare a bipartite quantum state and
send one part to Bob, and repeat this any number of times. Each time,
 they measure their respective parts, and communicate classically.
 Alice's task is to convince Bob that the state she can prepare is
 entangled. Bob (like \sch) accepts that QM describes the results
of the measurements he makes.
However Bob does not trust Alice.  
If the correlations between Bob's measurement results and the results Alice
reports {\em can} be explained by a LHS model for Bob then Bob will not be convinced that the
state is entangled; Alice could have
drawn a pure state at random from some ensemble and sent it to Bob,
and then chosen her result based on her knowledge of this LHS.
Conversely, if the correlations
{\em cannot} be so explained then the 
state must be entangled.
Therefore Alice will succeed in her task iff she can
create genuinely different ensembles for Bob, by {\em steering} Bob's state.



As EPR and \sch\ noted, steering may be demonstrated using any
pure entangled state, and the same is true of Bell-nonlocality
\cite{GisPLA91}.
But in the laboratory states are mixed.
In a seminal paper,   Werner
\cite{WerPRA89}  
asked the question:  can
all entangled states 
be used to demonstrate
Bell-nonlocality? As
Werner showed \cite{WerPRA89}, the surprising answer is: no --- a
hint of the complexity of bound entanglement \cite{HorEtalPRL98}
still being uncovered.

In this \paper, we address the following questions:  Can all
entangled states be used to demonstrate {\em steering}? Does a
demonstration of steering also demonstrate Bell-nonlocality?
We prove that in both cases the answer is again: no. Thus,
steerability is a distinct nonlocal property of some bipartite quantum states,
different from both Bell-nonlocality and nonseparability.

This \paper\ is structured as follows. We begin by finding the mathematical
formulation of the above operational definition of steering.
 From this it follows that steerability is stronger than
nonseparability, and weaker than Bell-nonlocality. We then show,
using two-qubit Werner states and Isotropic states, that this
is a strict hierarchy. Lastly, we consider Gaussian states with
Gaussian measurements. We determine the condition under which
steering can be demonstrated, and relate this to the Reid criterion
for the EPR ``paradox'' \cite{ReiPRA89}.

\section{Concepts of Quantum Nonlocality}
Let the set of all observables on the
Hilbert space for Alice's system be denoted ${\mathfrak D}_\alpha$.
We denote an element of ${\mathfrak D}_\alpha$ by $\hat{A}$, and
the set of eigenvalues $\cu{a}$ of $\hat{A}$ by $\lambda(\hat A)$.
By $P(a|\hat{A};W)$ we mean the probability that Alice will obtain
the result $a$ when she measures $\hat{A}$ on a system with state
matrix $W$. We denote the measurements that Alice is able to perform by
the set  ${\mathfrak M}_\alpha \subseteq {\mathfrak D}_\alpha$. (Note that, following Werner \cite{WerPRA89}, we are restricting to
projective measurements.)  The corresponding notations
for Bob, and for Alice and Bob jointly, are obvious. Thus, for example, \beq
P(a,b|\hat{A},\hat{B};W) = {\rm Tr}[(\hat\Pi_a^A\otimes
\hat\Pi_b^B) W], \eeq where $\hat\Pi_a^A$ is the
projector satisfying $\hat{A}\hat\Pi_a^A =
a\hat \Pi_a^A$.

The strongest sort of nonlocality in QM is
Bell-non\-locality \cite{BelPHY64}. This is exhibited in
an experiment on state $W$ iff the correlations between $a$ and $b$
cannot be explained by a LHV model. That is, if it is
{\em not} the case that for all $a\in \lambda(\hat A), b \in
\lambda(\hat B)$,  for all $\hat{A} \in {\mathfrak M}_\alpha,
\hat{B} \in {\mathfrak M}_\beta$, we have \beq \label{Bell-local1}
P(a,b| \hat A,\hat B;W) = \sum_\xi \wp(a|\hat A,\xi) \wp (b|\hat
B,\xi) \wp_\xi. \eeq Here, and below, $\wp(a|\hat A,\xi)$,
$\wp(b|\hat B,\xi)$ and $\wp_\xi$ denote some (positive,
normalized) probability distributions, involving the LHV $\xi$. We
say that a {\em state} is Bell-nonlocal iff there exists a measurement set
${\mathfrak M}_\alpha\times{\mathfrak M}_\beta$ that allows Bell-nonlocality
to be demonstrated.


A strictly weaker \cite{WerPRA89} concept is that of 
nonseparability or
entanglement. A nonseparable state is one that
{\em cannot} be written as $ W = \sum_\xi
{\sigma}_\xi \otimes \rho_\xi \,\wp_\xi$. Here, and below,
$\sigma_\xi \in {\mathfrak D}_\alpha$ and $\rho_\xi \in  {\mathfrak D}_\beta$ are
some (positive, normalized) quantum states. We can also give an operational
definition, by allowing  
Alice and Bob the ability to measure a quorum
of local observables, so that they can
reconstruct the state $W$ by tomography  \cite{DArEtalJPA01}. 
Thus a state $W$ is nonseparable 
iff it is {\em
not} the case that for all $a\in \lambda(\hat A), b \in
\lambda(\hat B)$, for all $\hat{A} \in {\mathfrak D}_\alpha,
\hat{B} \in {\mathfrak D}_\beta$, we have \beq \label{separable1}
P(a,b| \hat A,\hat B; W) = \sum_\xi P(a|\hat{A}; \sigma_\xi)
P(b|\hat{B};\rho_\xi) \wp_\xi. \eeq

Bell-nonlocality and nonseparability are both concepts that are
symmetric between Alice and Bob. However {\em steering}, \sch's term
for the 
EPR effect, is inherently asymmetric.
It is about whether Alice, by her choice of measurement $A$, can
collapse Bob's system into different types of states in the
different ensembles $E^A \equiv \cu{\tilde\rho^A_a:a\in
\lambda(A)}$. Here $\tilde{\rho}^A_a \equiv {\rm Tr}_\alpha[W (\hat\Pi_a^A\otimes {\bf I})] \in {\mathfrak D}_\beta$,
where the tilde denotes that this state is unnormalized (its norm is
the probability of its realization). Of course Alice
cannot affect Bob's unconditioned state
$\rho = {\rm Tr}_\alpha[W] = \sum_a \tilde \rho^A_a$
--- that would allow super-luminal signalling.
Nevertheless, as \sch\ said in 1935
\cite{SchPCP35}, ``It is rather discomforting that the theory should
allow a system to be steered \ldots into one or the other type 
of state at the experimenter's mercy in spite of [her] having no
access to it.'' As stated earlier, he was  ``not satisfied about  there being
enough experimental evidence for [it].'' 

The ``experimental evidence'' required 
by \sch\ is precisely that required
for Alice to succeed in the ``steering task'' defined in the introduction. 
The experiment can be
repeated at will, and we assume Bob's 
measurements enable him to do state tomography.  Prior to all
experiments, Bob demands that Alice announce the possible ensembles
$\cu{E^A:A \in {\mathfrak M}_\alpha}$ she can steer Bob's state into. In
any given run (after he has received his state), Bob should
 randomly pick an ensemble $E^A$, and ask Alice to prepare it.
Alice should then do so, by measuring $A$ on her
system, and announce to Bob the particular member
$\rho^A_a$ she has prepared. Over many runs, Bob can verify that
each state announced is indeed produced, and 
with the correct frequency ${\rm Tr}[\tilde \rho^A_a]$.

If Bob's system did have a pre-existing LHS $\rho_\xi$, 
then Alice could attempt to fool Bob, using her knowledge of $\xi$.
This state would be drawn at random from some prior ensemble of LHSs $F =
\cu{\wp_\xi\rho_\xi}$ with $\rho = \sum_\xi
\wp_\xi\rho_\xi$. Alice would then have to announce a LHS $\tilde{\rho}_a^A$
according to some stochastic map from $\xi$ to $a$. 
If, for all  $A \in {\mathfrak M}_\alpha$, and for all
$a\in \lambda(A)$, there exists a 
$\wp(a|A,\xi)$
such that
\beq \label{steering2}
\tilde{\rho}_a^A = \sum_\xi  \wp(a|A,\xi) \rho_\xi  \wp_\xi
\eeq
then Alice would have failed 
to convince Bob that she can steer his system. 
Conversely,  if Bob {\em cannot} find any ensemble $F$ and map $\wp(a|A,\xi)$ satisfying \erf{steering2} then Bob must admit that Alice can steer his system.

We can recast this definition as a `hybrid' of \erfand{Bell-local1}{separable1}: Alice's
 measurement strategy ${\mathfrak M}_\alpha$ on state $W$ exhibits steering iff
 it is {\em not} the case that
for all $a\in \lambda(\hat A), b \in \lambda(\hat B)$,  for all $\hat{A} \in {\mathfrak M}_\alpha, \hat{B} \in {\mathfrak D}_\beta$,  we have
\beq P(a,b| \hat A,\hat B; W) = \sum_\xi
\wp(a|\hat{A},\xi)P(b|\hat{B};\rho_\xi) \wp_\xi.
\label{Steer}\eeq
Iff there exists a measurement
strategy ${\mathfrak M}_\alpha$ that exhibits steering, we say that the state
$W$ is {\em steerable} (by Alice).

\begin{figure}
\begin{center}
\includegraphics[width=8.5cm]{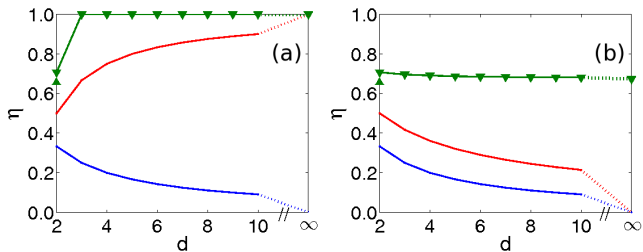}
\end{center}
\caption{(Color on-line.) Boundaries between classes of entangled
states for Werner (a) and Isotropic (b) states $W^\eta_d$. The
bottom (blue) line is $\eta_{\rm ent}$  (states are entangled iff 
$\eta > \eta_{\rm ent}$). The next (red) line is $\eta_{\rm steer}$, defined analogously
for steering. 
The top (green) line with down-arrows is an upper
bound on $\eta_{\rm Bell}$, defined analogously for Bell-nonlocality.  The up-arrows are lower bounds on $\eta_{\rm Bell}$
for $d=2$. The three classes are thus distinct. Dots join values at
finite $d$ with those at $d=\infty$.}
\label{combinedfigures}\end{figure}

Clearly steerability is stronger than nonseparability, but
Bell-nonlocality is stronger than steerability. At least one
of these relations must be ``strictly stronger'', because
of Werner's result \cite{WerPRA89}.
 In the following sections we prove   that {\em both} are ``strictly stronger''; 
 see Fig.~\ref{combinedfigures}. 

\section{Conditions for Steerability}
Below we derive necessary and sufficient conditions for steerability for three families of states. Crucial to the derivations is the concept of an \emph{optimal ensemble} $F^\star = \{\rho^\star_\xi \wp^\star_\xi\}$.  This is an ensemble such that if it cannot satisfy \erf{steering2} then no ensemble can satisfy it. In finding an optimal ensemble $F^\star$ we use the symmetries of $W$ and ${\mathfrak M}_\alpha$:
\begin{lemma} \label{lem_covariant}
Consider a  group $G$ with a unitary representation $U_{\alpha\beta}(g) = U_\alpha(g)\otimes U_\beta(g)$ on the Hilbert space for Alice and Bob. Say that $\forall A\in {\mathfrak M}_\alpha,\ \forall a\in \lambda(A),\ \forall g\in G,$  we have
$U_\alpha\dg(g)AU_\alpha(g) \in {\mathfrak M}_\alpha$ and
\beq
\tilde{\rho}^{U_\alpha\dg(g)AU_\alpha(g)}_a = U_\beta(g)\tilde{\rho}^{A}_aU_\beta\dg(g).
\eeq
Then there exists a $G$-covariant optimal ensemble:
$\forall g\in G,\ \{\rho^\star_\xi \wp^\star_\xi\} = \{U_\beta(g)\rho^\star_\xi U_\beta\dg(g) \wp^\star_\xi\}$.
\end{lemma}
\begin{proof}
Say there exists an ensemble $F=\{\rho_\xi\wp_\xi\}$ satisfying \erf{steering2}. Then under the conditions of lemma \ref{lem_covariant}, 
the $G$-covariant ensemble $F^\star = \{\rho_{(g,\xi)}^\star \wp_\xi d\mu_G(g) \}$, with $\rho_{(g,\xi)}^\star = U_\beta(g) \rho_\xi  U_\beta\dg(g)$, satisfies \erf{steering2} with the choice
$
\wp^\star(a|A,(g,\xi)) =  \wp(a|U_\alpha\dg(g)AU_\alpha(g),\xi)
$.
\end{proof}

\section{(i) Werner States} This family of states in $\mathbb{C}_d\otimes \mathbb{C}_d$ was introduced in Ref. \cite{WerPRA89}.
We parametrize it by $\eta\in\mathbb{R}$ such that $W^\eta_d$ is linear in $\eta$,  $W^\eta_d$ is a product state for $\eta=0$, and the largest permissible value for $\eta$ is 1:
 \beq
W_d^\eta=\left(\frac{d-1+\eta}{d-1}\right)\frac{\mathbf{I}}{d^2}-\left(\frac{\eta}{d-1}\right)\frac{\mathbf{V}}{d}.\label{WernerStates}\eeq
Here $\mathbf{I}$ is the identity and $\mathbf{V}$ the
``flip" operator ($\mathbf{V}\varphi\otimes\psi\equiv \psi\otimes\varphi$).
Werner states are nonseparable iff $\eta > \eta_{\rm ent} = 1/(d+1)$ \cite{WerPRA89}.
For $d=2$, the Werner states violate a Bell inequality if $\eta > 1/\sqrt{2}$  \cite{HorEtalPLA95}. This places an {\em upper} bound on $\eta_{\rm Bell}$, defined by  $W^\eta_d$ being Bell-nonlocal iff $\eta>\eta_{\rm Bell}$. For $d>2$ only the trivial upper bound of $1$ is known. However, Werner found a lower bound on $\eta_{\rm Bell}$ of $1-1/d$ \cite{WerPRA89}, which is strictly greater that $\eta_{\rm ent}$.

We now show that Werner's lower bound is in fact equal to $\eta_{\rm steer}$, defined by  $W^\eta_d$ being steerable iff $\eta>\eta_{\rm steer}$. We allow Alice all possible measurement strategies: ${\mathfrak M}_\alpha = {\mathfrak D}_\alpha$, and without loss of generality take the projectors to be rank-one: $\Pi_a^A = \ket{a}\bra{a}$. 
For Werner states, the conditions of lemma \ref{lem_covariant} are then satisfied for the $d$-dimensional unitary group ${\mathfrak U}(d)$. 
Specifically,  $g\to U$, and  $U_{\alpha\beta}(g) \to U\otimes U$ \cite{WerPRA89}. Again without loss of generality we can take the optimal ensemble to consist of pure states, in which case there is a unique 
covariant optimal ensemble, $F^\star =
\cu{\ket{\psi}\bra{\psi}d\mu_H({\psi})}$, where $d\mu_H(\psi)$ is the Haar measure over ${\mathfrak U}(d)$. 
Werner used the same construction; his LHVs for Bob
were in fact these LHSs. 

Now we determine when \erf{steering2} can be satisfied by this $F^\star$. Using $\tilde{\rho}_a^A=\langle a|W_d^\eta|a\rangle$ it is simple to show that, for
any $A \in {\mathfrak D}_\alpha$ and $a \in \lambda(A)$,
\beq \label{wernersteer}
\bra{a}\tilde{\rho}_a^A\ket{a}= (1-\eta)/d^2.
\eeq
Using the methods of Werner's proof we show that for any positive normalized distribution $\wp(a|A,\psi)$,
\beq \label{WernerGeometry}
\bra{a}\int d\mu_H(\psi) \ket{\psi}\bra{\psi} \wp(a|A,\psi) \ket{a} \geq 1/d^3.
\eeq
The upper bound is attained for the choice
\cite{WerPRA89}
\beq \wp(a|A,\psi)=\Big\{\begin{array}{ll}&1\ \ {\rm
if}\ |\langle\psi|a\rangle|\leq |\langle\psi|a'\rangle|,\ \forall
a'\neq a\\
&0\  \ {\rm otherwise.} \end{array}\Big.\label{WernerMethod} \eeq
Comparing this  with \erf{wernersteer} we see that steering can be demonstrated if
$(1-\eta)/d^2 < 1/ d^3$. Moreover, it is easy to verify that when this inequality is saturated, \erf{WernerMethod} satisfies \erf{steering2}. Thus $\eta_{\rm steer} = 1-{1}/{d}$.

Recently a new lower bound for $\eta_{\rm Bell}$ was found for $d=2$ \cite{AciEtalPRA06}, greater than $\eta_{\rm steer}$, as shown in Fig.~\ref{combinedfigures}. 
This proves that steerability is strictly weaker than Bell-nonlocality as well as being strictly stronger than non-separability.

\section{(ii) Isotropic States} This family, introduced in \cite{HorHorPRA99}, can be parametrized identically to the Werner states:
\beq W_d^\eta=(1-\eta){\mathbf{I}}/{d^2}+\eta \mathbf{P}_+,
\label{isotropicstates}\eeq where
$\mathbf{P}_+=\ket{\psi_+}\bra{\psi_+}$, where $\ket{\psi_+} =\sum_{i=1}^{d}\ket{i}\ket{i}/\sqrt{d}$. For $d=2$ the Isotropic states are identical to Werner states.
Isotropic states are nonseparable iff  $\eta > \eta_{\rm ent} = 1/(d+1)$ \cite{HorHorPRA99}.
A non-trivial upper bound on $\eta_{\rm Bell}$ for all $d$ is known \cite{ColEtalPRL02};  Ref.~\cite{AciEtalPRA06} gives a lower bound for $d=2$.

To determine $\eta_{\rm steer}$ for isotropic states, we follow the same method as for
Werner states, except that this time $U_{\alpha\beta} = U^*\otimes U$ \cite{HorHorPRA99}. Instead of \erf{wernersteer} we obtain
\beq \label{isosteer}
\bra{a}\tilde{\rho}_a^A \ket{a}= \xfrac{\eta}{d} + \xfrac{(1-\eta)}{d^2},
\eeq
and instead of \erf{WernerGeometry},
we show that 
\beq \int d\mu_H(\psi) \st{ \langle{a}|{\psi}\rangle }^2
\wp(a|A,\psi) \leq { H_d/d^2 }, \eeq  where
$H_d=\sum_{n=1}^{d}(1/n)$ is the Harmonic series. 
The upper bound
is attained for the choice \beq
\wp(a|A,\psi)=\Big\{\begin{array}{ll}&1\ \ {\rm if}\
|\langle\psi|a\rangle|\geq |\langle\psi|a'\rangle|,\ \forall
a'\neq a\\
&0\  \ {\rm otherwise.} \end{array}\Big.\label{IsoMethod} \eeq
Comparing this result with \erf{isosteer}, we see that isotropic states are steerable if
$\eta >   \ro{ H_d-1}/(d-1)$.
Moreover, it is easy to verify that when this inequality is saturated, \erf{IsoMethod} satisfies \erf{steering2}. Thus $\eta_{\rm steer}=\ro{ H_d-1}/(d-1)$.

\section{(iii) Gaussian States}
Finally we investigate a general (multimode) bipartite Gaussian state $W$ \cite{GieCirPRA02}.  Such a state may be defined by its covariance matrix (CM) $V_{\alpha\beta}$. In (Alice, Bob) block form it appears as
\beq \label{cov-matrix}
{\rm CM}[W] =V_{\alpha\beta}=\left(\begin{matrix}
V_\alpha &   C \\
C\tp  & V_\beta
\end{matrix} \right).
\eeq
This represents a valid state iff $V_{\alpha\beta}+i\Sigma_{\alpha\beta}\geq 0$ \cite{GieCirPRA02}. This is a linear matrix inequality (LMI), in which  $\Sigma_{\alpha\beta} = \Sigma_\alpha \oplus \Sigma_\beta$ is a symplectic matrix proportional to $\hbar$.

Rather than addressing steerability in general, we consider the case where Alice can only make  Gaussian measurements, denoted by $\mathfrak{G}_\alpha$. A measurement $A \in \mathfrak{G}_\alpha$ is described by a Gaussian positive operator with a CM $T^A$ satisfying $T^A + i\Sigma_\alpha \geq 0$ \cite{GieCirPRA02}. When Alice makes such a measurement, Bob's conditioned state $\rho_a^A$  is Gaussian with a CM
$V_\beta^A =  V_\beta -  C(T^A + V_\alpha)^{-1}C\tp$ \cite{ZhoEtalORC96}.
\begin{theorem}
The Gaussian state $W$ defined in \erf{cov-matrix} is {\em not} steerable by Alice's Gaussian measurements iff
\beq \label{LMIL}
V_{\alpha\beta} + {\bf 0}_\alpha\oplus i\Sigma_\beta \geq 0.
\eeq
\end{theorem}
\begin{proof} The proof has two parts.
First, suppose \erf{LMIL} is true. Then using matrix inversion formulas \cite{ZhoEtalORC96}, one sees that the matrix $U \equiv V_\beta -  CV_\alpha^{-1}C\tp$
satisfies
\beq
U+i\Sigma_\beta \geq 0 \ \textrm{ and }\  \forall A \in \mathfrak{G}_\alpha,\
V_\beta^A - U \geq 0 . \label{definE}
\eeq
  The first LMI allows us to define an ensemble $F^U=\{\rho_\xi^U\wp^U_\xi\}$ of Gaussian states with  CM$[\rho^U_\xi]=U$,  distinguished by their mean vectors ($\xi$). The second LMI implies that $\forall A,\ \rho^A_a$ is a Gaussian mixture (over $\xi$) of such states. 
Therefore $W$ is not steerable by Alice. 

Now suppose $W$ is not steerable. Then there is some ensemble $F$ satisfying \erf{steering2}. From the fact that $V_\beta^A$ is  independent of $a$, one sees that
$U = \sum_\xi \wp_\xi \times {\rm CM}[\rho_\xi]$ satisfies \erf{definE}. But unless (\ref{LMIL}) is true, one sees that no such $U$ satsifying \erf{definE} exists (again using standard matrix analysis \cite{ZhoEtalORC96}).
Therefore (\ref{LMIL}) must be true.
\end{proof}

For the simplest case where Alice and Bob each have one mode with correlated positions $q$ and momenta $p$,  Reid \cite{ReiPRA89} has argued the  EPR ``paradox'' is demonstrated iff the product of the conditional variances $V(q_\beta|q_\alpha)$ and
$V(p_\beta|p_\alpha)$  violates the uncertainty principle. It is easy to verify that this occurs under
precisely the same conditions as when \erf{LMIL} is violated. This confirms
that the EPR ``paradox'' is merely a particular case of steering. As is well known \cite{BowEtalPRA04}, the Reid conditions are strictly stronger than the conditions for nonseparability.

We conclude with a brief listing of open questions. First, 
are there asymmetric states that are steerable by Alice
but not by Bob? Second, Bell-nonlocality is necessary and sufficient for certain tasks \cite{BruEtalPRL04}, and likewise nonseparability \cite{MasPRL06}. Is there
a  task (beyond the defining one) for which steerability is similarly useful?
Third, do there exist steering analogs of Bell-operators and entanglement witnesses? Finally, we note that we expect many applications of the concept of steering in quantum measurement theory and experimental quantum information.

\acknowledgments

This work was supported by the ARC and the  State of Queensland.
We thank Rob Spekkens, Volkher Scholz, Antonio Acin, and Michael Hall for useful discussions. 

\bibliography{Steering}

\begin{thebibliography}{23}
\expandafter\ifx\csname natexlab\endcsname\relax\def\natexlab#1{#1}\fi
\expandafter\ifx\csname bibnamefont\endcsname\relax
  \def\bibnamefont#1{#1}\fi
\expandafter\ifx\csname bibfnamefont\endcsname\relax
  \def\bibfnamefont#1{#1}\fi
\expandafter\ifx\csname citenamefont\endcsname\relax
  \def\citenamefont#1{#1}\fi
\expandafter\ifx\csname url\endcsname\relax
  \def\url#1{\texttt{#1}}\fi
\expandafter\ifx\csname urlprefix\endcsname\relax\def\urlprefix{URL }\fi
\providecommand{\bibinfo}[2]{#2}
\providecommand{\eprint}[2][]{\url{#2}}

\bibitem[{\citenamefont{Einstein et~al.}(1935)\citenamefont{Einstein, Podolsky,
  and Rosen}}]{EinEtalPR35}
\bibinfo{author}{\bibfnamefont{A.}~\bibnamefont{Einstein}},
  \bibinfo{author}{\bibfnamefont{B.}~\bibnamefont{Podolsky}}, \bibnamefont{and}
  \bibinfo{author}{\bibfnamefont{N.}~\bibnamefont{Rosen}},
  \bibinfo{journal}{Phys. Rev.} \textbf{\bibinfo{volume}{47}},
  \bibinfo{pages}{777} (\bibinfo{year}{1935}).

\bibitem[{\citenamefont{Schr\"odinger}(1935)}]{SchPCP35}
\bibinfo{author}{\bibfnamefont{E.}~\bibnamefont{Schr\"odinger}},
  \bibinfo{journal}{Proc. Camb. Phil. Soc.} \textbf{\bibinfo{volume}{31}},
  \bibinfo{pages}{553} (\bibinfo{year}{1935}), \bibinfo{note}{\emph{ibid.} {\bf
  32}, 446, (1936).}

\bibitem[{\citenamefont{Schr\"odinger}(1926)}]{SchAdP26}
\bibinfo{author}{\bibfnamefont{E.}~\bibnamefont{Schr\"odinger}},
  \bibinfo{journal}{Ann. der Phys.} \textbf{\bibinfo{volume}{79}},
  \bibinfo{pages}{361} (\bibinfo{year}{1926}), \bibinfo{note}{\emph{ibid.} {\bf
  79}, 489 (1926); \emph{ibid.} {\bf 80}, 437 (1926); \emph{ibid.} {\bf 81},
  109 (1926).}

\bibitem[{\citenamefont{Vuji\'{c}i\v{c} and Herbut}(1988)}]{VujHerJPA88}
\bibinfo{author}{\bibfnamefont{M.}~\bibnamefont{Vuji\'{c}i\v{c}}}
  \bibnamefont{and} \bibinfo{author}{\bibfnamefont{F.}~\bibnamefont{Herbut}},
  \bibinfo{journal}{J. Phys. A} \textbf{\bibinfo{volume}{21}},
  \bibinfo{pages}{2931} (\bibinfo{year}{1988}).

\bibitem[{\citenamefont{Verstraete}(2002)}]{VerPhD02}
\bibinfo{author}{\bibfnamefont{F.}~\bibnamefont{Verstraete}}, Ph.D. thesis,
  \bibinfo{school}{Katholieke University Leuven} (\bibinfo{year}{2002}).

\bibitem[{\citenamefont{Clifton et~al.}(2003)\citenamefont{Clifton, Bub, and
  Halvorson}}]{CliEtalFoP03}
\bibinfo{author}{\bibfnamefont{R.}~\bibnamefont{Clifton}},
  \bibinfo{author}{\bibfnamefont{J.}~\bibnamefont{Bub}}, \bibnamefont{and}
  \bibinfo{author}{\bibfnamefont{H.}~\bibnamefont{Halvorson}},
  \bibinfo{journal}{Found. Phys.} \textbf{\bibinfo{volume}{33}},
  \bibinfo{pages}{1561} (\bibinfo{year}{2003}).

\bibitem[{\citenamefont{Spekkens}(2007)}]{SpePRA07}
\bibinfo{author}{\bibfnamefont{R.~W.} \bibnamefont{Spekkens}},
  \bibinfo{journal}{Phys. Rev. A} \textbf{\bibinfo{volume}{75}},
  \bibinfo{pages}{032110} (\bibinfo{year}{2007}).

\bibitem[{\citenamefont{Kirkpatrick}(2006)}]{KirFPL06}
\bibinfo{author}{\bibfnamefont{K.~A.} \bibnamefont{Kirkpatrick}},
  \bibinfo{journal}{Found. Phys. Lett} \textbf{\bibinfo{volume}{19}},
  \bibinfo{pages}{95} (\bibinfo{year}{2006}).

\bibitem[{\citenamefont{Bell}(1964)}]{BelPHY64}
\bibinfo{author}{\bibfnamefont{J.~S.} \bibnamefont{Bell}},
  \bibinfo{journal}{Physics (Long Island City, N.Y.)}
  \textbf{\bibinfo{volume}{1}}, \bibinfo{pages}{195} (\bibinfo{year}{1964}).

\bibitem[{\citenamefont{Gisin}(1991)}]{GisPLA91}
\bibinfo{author}{\bibfnamefont{N.}~\bibnamefont{Gisin}},
  \bibinfo{journal}{Phys. Lett. A} \textbf{\bibinfo{volume}{154}},
  \bibinfo{pages}{201} (\bibinfo{year}{1991}).

\bibitem[{\citenamefont{Werner}(1989)}]{WerPRA89}
\bibinfo{author}{\bibfnamefont{R.~F.} \bibnamefont{Werner}},
  \bibinfo{journal}{Phys. Rev. A} \textbf{\bibinfo{volume}{40}},
  \bibinfo{pages}{4277} (\bibinfo{year}{1989}).

\bibitem[{\citenamefont{Horodecki et~al.}(1998)\citenamefont{Horodecki,
  Horodecki, and Horodecki}}]{HorEtalPRL98}
\bibinfo{author}{\bibfnamefont{M.}~\bibnamefont{Horodecki}},
  \bibinfo{author}{\bibfnamefont{P.}~\bibnamefont{Horodecki}},
  \bibnamefont{and}
  \bibinfo{author}{\bibfnamefont{R.}~\bibnamefont{Horodecki}},
  \bibinfo{journal}{Phys. Rev. Lett.} \textbf{\bibinfo{volume}{80}},
  \bibinfo{pages}{5239} (\bibinfo{year}{1998}).

\bibitem[{\citenamefont{Reid}(1989)}]{ReiPRA89}
\bibinfo{author}{\bibfnamefont{M.~D.} \bibnamefont{Reid}},
  \bibinfo{journal}{Phys. Rev. A} \textbf{\bibinfo{volume}{40}},
  \bibinfo{pages}{913} (\bibinfo{year}{1989}).

\bibitem[{\citenamefont{D'Ariano et~al.}(2001)\citenamefont{D'Ariano, Maccone,
  and Paris}}]{DArEtalJPA01}
\bibinfo{author}{\bibfnamefont{G.~M.} \bibnamefont{D'Ariano}},
  \bibinfo{author}{\bibfnamefont{L.}~\bibnamefont{Maccone}}, \bibnamefont{and}
  \bibinfo{author}{\bibfnamefont{M.~G.~A.} \bibnamefont{Paris}},
  \bibinfo{journal}{J. Phys. A} \textbf{\bibinfo{volume}{34}},
  \bibinfo{pages}{93} (\bibinfo{year}{2001}).

\bibitem[{\citenamefont{Horodecki et~al.}(1995)\citenamefont{Horodecki,
  Horodecki, and Horodecki}}]{HorEtalPLA95}
\bibinfo{author}{\bibfnamefont{R.}~\bibnamefont{Horodecki}},
  \bibinfo{author}{\bibfnamefont{P.}~\bibnamefont{Horodecki}},
  \bibnamefont{and}
  \bibinfo{author}{\bibfnamefont{M.}~\bibnamefont{Horodecki}},
  \bibinfo{journal}{Phys. Lett. A} \textbf{\bibinfo{volume}{200}},
  \bibinfo{pages}{340} (\bibinfo{year}{1995}).

\bibitem[{\citenamefont{Ac\`{i}n et~al.}(2006)\citenamefont{Ac\`{i}n, Gisin,
  and Toner}}]{AciEtalPRA06}
\bibinfo{author}{\bibfnamefont{A.}~\bibnamefont{Ac\`{i}n}},
  \bibinfo{author}{\bibfnamefont{N.}~\bibnamefont{Gisin}}, \bibnamefont{and}
  \bibinfo{author}{\bibfnamefont{B.}~\bibnamefont{Toner}},
  \bibinfo{journal}{Phys. Rev. A} \textbf{\bibinfo{volume}{73}},
  \bibinfo{pages}{062105} (\bibinfo{year}{2006}).

\bibitem[{\citenamefont{Horodecki and Horodecki}(1999)}]{HorHorPRA99}
\bibinfo{author}{\bibfnamefont{M.}~\bibnamefont{Horodecki}} \bibnamefont{and}
  \bibinfo{author}{\bibfnamefont{P.}~\bibnamefont{Horodecki}},
  \bibinfo{journal}{Phys. Rev. A} \textbf{\bibinfo{volume}{59}},
  \bibinfo{pages}{4206} (\bibinfo{year}{1999}).

\bibitem[{\citenamefont{Collins}(2002)}]{ColEtalPRL02}
\bibinfo{author}{\bibfnamefont{D.}~\bibnamefont{Collins}},
  \bibinfo{journal}{\emph{et al.}, Phys. Rev. Lett.}
  \textbf{\bibinfo{volume}{88}}, \bibinfo{pages}{040404}
  (\bibinfo{year}{2002}).

\bibitem[{\citenamefont{Giedke and Cirac}(2002)}]{GieCirPRA02}
\bibinfo{author}{\bibfnamefont{G.}~\bibnamefont{Giedke}} \bibnamefont{and}
  \bibinfo{author}{\bibfnamefont{J.}~\bibnamefont{Cirac}},
  \bibinfo{journal}{Phys. Rev. A} \textbf{\bibinfo{volume}{66}},
  \bibinfo{pages}{032316} (\bibinfo{year}{2002}).

\bibitem[{\citenamefont{Zhou et~al.}(1996)\citenamefont{Zhou, Doyle, and
  Glover}}]{ZhoEtalORC96}
\bibinfo{author}{\bibfnamefont{K.}~\bibnamefont{Zhou}},
  \bibinfo{author}{\bibfnamefont{J.~C.} \bibnamefont{Doyle}}, \bibnamefont{and}
  \bibinfo{author}{\bibfnamefont{K.}~\bibnamefont{Glover}},
  \emph{\bibinfo{title}{Optimal and Robust Control}}
  (\bibinfo{publisher}{Prentice-Hall, Englewood Cliffs, NJ},
  \bibinfo{year}{1996}).

\bibitem[{\citenamefont{Bowen}(2004)}]{BowEtalPRA04}
\bibinfo{author}{\bibfnamefont{W.~P.} \bibnamefont{Bowen}},
  \bibinfo{journal}{\emph{et al}, Phys. Rev. A} \textbf{\bibinfo{volume}{69}},
  \bibinfo{pages}{012304} (\bibinfo{year}{2004}).

\bibitem[{\citenamefont{Brukner}(2004)}]{BruEtalPRL04}
\bibinfo{author}{\bibfnamefont{C.}~\bibnamefont{Brukner}},
  \bibinfo{journal}{\emph{et al.}, Phys. Rev. Lett.}
  \textbf{\bibinfo{volume}{92}}, \bibinfo{pages}{127901}
  (\bibinfo{year}{2004}).

\bibitem[{\citenamefont{Masanes}(2006)}]{MasPRL06}
\bibinfo{author}{\bibfnamefont{L.}~\bibnamefont{Masanes}},
  \bibinfo{journal}{Phys. Rev. Lett.} \textbf{\bibinfo{volume}{96}},
  \bibinfo{pages}{150501} (\bibinfo{year}{2006}).

\end{thebibliography}

\end{document}